\documentstyle[preprint,aps,amsfonts,prd,epsf]{revtex}

\begin{document}

\title {Renormalization Group Approach to the Dynamical Casimir Effect}
    
\author{Diego A.\ R.\ Dalvit  \thanks{dalvit@df.uba.ar} and
Francisco D.\ Mazzitelli \thanks{fmazzi@df.uba.ar}}

\address{ {\it
Departamento de F\'\i sica, Facultad de Ciencias Exactas y Naturales\\ 
Universidad de Buenos Aires- Ciudad Universitaria, Pabell\' on I\\ 
1428 Buenos Aires, Argentina}}

\maketitle

\begin{abstract}
In this paper we study
the one dimensional dynamical Casimir effect. We consider a one dimensional
cavity formed by two mirrors, one
of which performs an oscillatory motion with a frequency  resonant with
the cavity. 
The naive solution, perturbative in powers of the amplitude, contains
secular terms. Therefore it is valid only in the short time
limit.
Using a renormalization group technique to resum these
terms, we obtain an improved analytical solution
which is valid for longer times. We
discuss the generation of peaks in the density energy profile and show
that the total energy inside the cavity increases exponentially. 
\end{abstract}


\section{INTRODUCTION}

The problem of quantum fluctuations inside cavities has attracted the
attention since a long time ago \cite{Casimir}. 
A way to study the structure of the vacuum is
to distort it by changing the configuration of the cavity in time 
\cite{Moore,FD}. The 
simplest configuration is that of a one dimensional cavity formed
by two perfectly reflecting mirrors, one of which is fixed and the other
one is allowed to move in a predetermined way, or rather, its motion is
determined by the backreaction of the electromagnetic field \cite{Rafael}.
There are
in fact only a few predetermined motions which allow an exact resolution
of the problem. In \cite{Law} a special motion for the mirror, 
which has an 
exact solution, has been considered, and it has been shown
that the Casimir force may be resonantly enhanced. In \cite{Cole}
a geometrical method for solving the problem for arbitrary
wall motions has been developed, and basically the same structure for the
electromagnetic field within the cavity has been found. 
Of special interest are the cases 
where the moving mirror oscillates with one of the eigenfrequencies of the
unperturbed cavity \cite{DodPLA,DodJMP,DodPRA}.
A naive approach is to make perturbations in the
amplitude of oscillation. However, this perturbative treatment has
only a very limited range of validity: the appearance of secular terms
proportional to the time imply that after a short period the approximation
breaks down. In this paper we apply a method inspired 
in the Renormalization
Group (RG) to treat these singular perturbations. The method has a wide
range of application \cite{goldenfeld}, 
especially to ordinary differential equation
problems involving multiple scales, boundary layers, asymptotic matching
and WKB analysis. The main advantage of the RG method is to provide a
simple and unified calculational method for all problems of this sort.
In our present case, the application of the method to the dynamical Casimir
effect permits to get a solution for the structure of the electromagnetic
field within the cavities that is valid for a period of time longer than
that of the
perturbative case. With the solution at hand, we study local properties
of the field such as the energy density. In agreement with other
authors \cite{Cole}, we show that the resonant moving wall induces 
an exponential
growth of the total energy, and that peaks form inside the cavity, which
travel at the speed of light bouncing against the walls.

We consider a one dimensional cavity formed by two perfectly reflecting 
mirrors. One of them is fixed at $x=0$, while the other one performs
an oscillatory motion $L(t)=L_0 [1+\epsilon \sin (\frac{q \pi t}{L_0})]$,
with $\epsilon \ll 1$ and 
$q \in \Bbb N$, i.e., the moving mirror oscillates with a frequency equal
to one of the eigenfrequencies of the cavity.
We shall assume that the oscillations begin at $t=0$, and that the
mirror is at rest for $t <0 $. Note that we shall not treat the moving
mirror as a degree of freedom (either classical or quantum), but just
as a given time-dependent boundary for the electromagnetic field
inside the cavity.
The vector potential $A(x,t)$  satisfies the one dimensional field equation 
 $\Box A=0$  and the boundary conditions  $A(x=0,t)=A(L(t),t)=0$ for all 
times. Therefore one can express the field inside the cavity as
\begin{equation}
A(x,t)= \sum_{k=1}^{\infty} \left[ a_k \psi_k(x,t) + a_k^{\dagger} 
                                   \psi_k^*(x,t) \right] ,
\end{equation}
where the mode functions $\psi_k(x,t)$ are positive frequency modes
for $t<0$, and $a_k$ and $a_k^{\dagger}$ are time-independent
annihilation and creation operators, respectively. 

If one writes
the modes in terms of a function $R(t)$ as \cite{Moore} 
\begin{equation}
\psi_k(x,t)=\frac{i}{\sqrt{4 \pi k}} 
\left( e^{-i k \pi R(t+x)} - e^{-i k \pi R(t-x)} \right) ,
\end{equation}
the boundary  conditions are met provided that 
\begin{equation}
R(t+L(t))-R(t-L(t))=2 .
\label{mooreq}
\end{equation}
The complete solution to the problem involves finding a solution $R(t)$
in terms of the prescribed motion $L(t)$. The modes are positive
frequency modes for $t<0$ if $R(t)={t/L_0}$ for $- L_0 \leq t \leq L_0$, 
which is indeed a solution to Eq.(\ref{mooreq}) for $t<0$.
Note that the boundary condition for $R(t)$ involves its values
over the whole range of time $- L_0 \leq t \leq L_0$. 

In what follows, we will
describe a method to find an  analytical approximation to the
solution of Eq.(\ref{mooreq}). We shall first obtain a perturbative
solution by expanding in powers of the amplitude $\epsilon$
(Section 3). As
this perturbative solution will contain secular terms, it will
be valid only for short times, i.e. $\epsilon \, \frac{t}{L_0}< 1$. Using 
RG techniques, we will be able to perform
a resummation of the secular terms and obtain an analytical
approximation valid for a longer period of time $\epsilon^2 \frac{t}{L_0}< 1$
(Section 4). In Section 5 we will use this solution to
describe the evolution
of the mean value of the energy density inside the cavity.
In order to get acquainted with the renormalization
group method, in Section 2 we will describe a simple example where
the resummation of the secular terms is performed for a particular
ordinary differential equation. The reader already familiarized 
with the work of Ref. \cite{goldenfeld} can skip the next Section.

\section{A SIMPLE EXAMPLE}

Let us consider the  Rayleigh equation
\begin{equation}
{d^2y\over dt^2}+y +\epsilon \left\{{1\over3}\left ({dy\over dt}\right)
^3 - {dy\over dt}\right\}=0 ,
\label{rayl}
\end{equation}
where $\epsilon$ is a small number. This is an interesting 
oscillator because it
can be shown that, for any initial condition and any positive
$\epsilon$, the exact solution becomes periodic at long times and 
therefore approaches a limit circle in phase space \cite{bender}.  

The Rayleigh equation can be solved  perturbatively
using an
expansion in powers of $\epsilon$, that is,
$y=y_0+\epsilon y_1 + {\cal O}(\epsilon^2)$.
Up to first order in $\epsilon$,
the perturbative solution reads
\begin{eqnarray}
y(t) &=& Y_0 \sin (t+\Theta_0) + \epsilon\left\{ {Y_0\over 2}\left (
1-{Y_0^2\over 4}\right )(t-t_0) \sin (t+\Theta_0)+\right.\nonumber\\
&& \left . {Y_0^3\over 96} \left[ \cos (3(t+\Theta_0)) - \cos (t+\Theta_0)
\right] \right \} + {\cal O}(\epsilon ^2) ,
\label{pertt0}
\end{eqnarray}
where $Y_0$ and $\Theta_0$ are constants determined by the 
initial conditions at arbitrary $t=t_0$. This perturbative
solution does not become periodic and, therefore,
it is not a good approximation for long times. Indeed, due to the presence of
the secular term, the naive perturbative solution is valid
only for  times close to the initial time
$t_0$, and breaks down for $\epsilon \, (t-t_0) \geq 1$. 
This is typical of systems showing parametric 
resonance. Usually one
has a system weakly coupled to an external resonant force, and as one
tries to make a perturbative analysis, the corrections possess secular terms,
i.e. terms that grow linearly with time. In the following we shall adopt
the RG method to treat singular perturbations and we shall
use it to make an improvement to the perturbative solution, which shall
be valid for longer times, $\epsilon^2 (t-t_0) <1$. The basic idea 
\cite{goldenfeld} is to
introduce an arbitrary time $\tau$, split $t-t_0$ as $(t-\tau) +\tau - t_0$, 
and absorb the terms proportional to  $\tau - t_0$ into the 
`renormalized' counterparts
 $Y(\tau)$ and $\Theta(\tau)$ of the `bare' parameters 
contained in the zeroth order solution, that is, $Y_0$ and $\Theta_0$.
Using this idea one eliminates the secular terms proportional to $\tau
-t_0$,
and the function $y(t)$ takes the form
\begin{eqnarray}
y(t) &=& Y \sin (t+\Theta) + \epsilon\left\{ {Y\over 2}\left (
1-{Y\over 4}\right )(t-\tau ) \sin (t+\Theta)+ \right .\nonumber\\
&& \left . {Y^3\over 96} \left[ \cos (3(t+\Theta )) - \cos (t+\Theta)
\right] \right \} + {\cal O}(\epsilon ^2) ,
\label{perttau}
\end{eqnarray}
where now $Y$ and $\Theta$ are functions of $\tau$. 
Now comes the crucial point. 
As $\tau$ does not appear in the original
equation nor in the initial conditions, the solution
Eq.(\ref {perttau}) should not depend on $\tau$. Therefore,
the partial derivative with respect to $\tau$ should
vanish, i.e., $\left (\partial y/\partial
\tau\right )_t=0$ for any $t$. This is the RG equation,
which implies
\begin{eqnarray}
{dY\over d\tau}&=&\epsilon {Y\over 2}\left (1-{Y^2\over 4}\right
) +{\cal O}(\epsilon^2) , \nonumber \\
 {d\Theta\over d \tau} &=& {\cal O}(\epsilon^2) .
\label{RGE}
\end{eqnarray}
The solutions to these equations are
\begin{eqnarray}
Y(\tau ) &=& Y(t_0) \left[ e^{-\epsilon (\tau-t_0) } +
{Y^2(t_0) \over 4}(1-e^{-\epsilon 
(\tau -t_0)})
\right]^{-1/2}+{\cal O}(\epsilon ^2 (\tau - t_0)) , \nonumber \\
\Theta (\tau) &=& \Theta(t_0) + {\cal O}(\epsilon^2(\tau -t_0)) ,
\label{rtau}
\end{eqnarray}
where $Y(t_0)$ and $\Theta(t_0)$ are constants to be determined by the initial
conditions.
We still have the freedom to choose the arbitrary time $\tau$.
The obvious choice is $t=\tau$, since in this way the secular term 
proportional to $t-\tau$ in Eq. (\ref{perttau})
disappears. Assuming the initial condition 
$y(t_0)=0$, $\dot{y}(t_0)=2 a$, with $a$ any real number, we find
$Y(t_0)=2 a$ and $\Theta(t_0)=-t_0$. Finally,
the RG improved solution reads
\begin{equation}
y(t)=Y(t) \sin (t-t_0)+\epsilon {Y(t)^3\over 96}
[\cos (3 (t-t_0))- \cos (t-t_0)] + {\cal O}(\epsilon ^2)  ,
\label{improvsol}
\end{equation}
that is valid for $\epsilon^2 (t-t_0) < 1$. Note that the 
improved solution
becomes periodic  and approaches a limit circle of
radius $2$ for $\epsilon (t-t_0) \gg 1$. 

It is interesting to remark the analogy with the usual RG approach in 
quantum field
theory: $t_0$ plays the role of the ultraviolet cutoff
(although there are no divergences  but secular terms here), 
$Y_0$ and $\Theta_0$
are the bare coupling constants, and $Y$ and $\Theta$ are 
the renormalized counterparts evaluated at the 'scale'
$\tau$. The equation is  'renormalizable' because the
secular terms can be absorbed into the bare parameters.
As anticipated, the RG  is a straightforward method
to obtain, from the naive perturbative solution, an improved
solution which is valid for longer times. In this particular example,
it can be shown that the RG method is equivalent to multiple-scale
analysis  \cite{bender}, with the additional practical advantadge 
that it is
not necessary to know a priori the multiple time scales.
 
\section{PERTURBATIVE SOLUTION TO THE DYNAMICAL CASIMIR EFFECT}

We will now solve Eq. (\ref{mooreq}) using a naive perturbative expansion. 
We expand 
the function $R(t)$ in terms of the small amplitude $\epsilon$ 
and retain first order terms only, $R(t)=R_0(t) + \epsilon R_1(t)$.
Equating terms of the same order 
we get
\begin{eqnarray}
R_0(t+L_0) - R_0(t-L_0) &=& 2 \label{Rcero} , \\
R_1(t+L_0) - R_1(t-L_0) &=& -L_0 \sin(\frac{q \pi t}{L_0}) 
[R'_0(t+L_0)+R'_0(t-L_0)] .
\label{Runo}
\end{eqnarray}
The general solution to Eq.(\ref{Rcero}) is
\begin{equation}
R_0(t) = a + \frac{t}{L_0} + \sum_{n \geq 1} \left[
A_n \sin(\frac{n \pi t}{L_0}) + B_n \cos(\frac{n \pi t}{L_0})
\right] ,
\label{r0}
\end{equation}
where $a, A_n$ and $B_n$ are constants determined by the boundary condition,
that is, by the value of $R(t)$ for $-L_0 \leq t \leq L_0$. Introducing this 
solution into Eq. (\ref{Runo}) we 
obtain
\begin{eqnarray}
-\frac{1}{2} \left[R_1(t+L_0)-R_1(t-L_0)\right] &=&
\sin(\frac{q \pi t}{L_0}) + \frac{\pi}{2} \sum_{n \geq 1} n (-1)^n \times 
\nonumber\\
&& \left\{
A_n \left[ \sin \left( \frac{(q+n) \pi t}{L_0} \right) + 
\sin \left( \frac{(q-n) \pi t}{L_0} \right)
    \right] + 
\right. \nonumber \\
&& \left. 
~~ B_n \left[ \cos \left( \frac{(q+n) \pi t}{L_0} \right) - 
\cos \left( \frac{(q-n) \pi t}{L_0} \right)
    \right] 
\right\} ,
\end{eqnarray}
whose general solution reads

\begin{eqnarray}
R_1(t) &=& (-1)^{q+1} \frac{t}{L_0} 
\left\{
\sin(\frac{q \pi t}{L_0}) + \frac{\pi}{2} \sum_{n \geq 1} n  \times 
\left(
A_n \left[ \sin \left( \frac{(q+n) \pi t}{L_0} \right) + 
\sin \left( \frac{(q-n) \pi t}{L_0} \right)
    \right] 
\right. +
\right. 
\nonumber \\
&& 
\left.
\left.
B_n \left[ \cos \left( \frac{(q+n) \pi t}{L_0} \right) - 
\cos \left( \frac{(q-n) \pi t}{L_0} \right)
    \right]
\right)
\right\} + g(t) ,
\label{r1}
\end{eqnarray}
where $g(t)$ is an arbitrary periodic function with period $2 L_0$.
We see that, as in the
case of the Rayleigh oscillator, the perturbative correction contains
secular terms that grow linearly in time. Therefore, this approximation
will be valid only for short times, that is, $\epsilon \, \frac{t}{L_0} <1$. 

If we assume that the
boundary condition for $R(t)$ is already satisfied by $R_0(t)$, 
then the
periodic function $g(t)$ must be chosen in such a way that
$R_1(t)=0$ for $-L_0 \leq t \leq L_0$. Therefore
\begin{eqnarray}
g(2pL_0+z) &=& (-1)^{q} \frac{z}{L_0} 
\left\{
\sin(\frac{q \pi z}{L_0}) + \frac{\pi}{2} \sum_{n \geq 1} n  \times 
\right.
\nonumber \\
&&
\left(
A_n \left[ \sin \left( \frac{(q+n) \pi z}{L_0} \right) +
 \sin \left( \frac{(q-n) \pi z}{L_0} \right) \right] +
\right.
\nonumber \\
&&
\left.
\left. 
~~~ B_n \left[ \cos \left( \frac{(q+n) \pi z}{L_0} \right) - 
\cos \left( \frac{(q-n) \pi z}{L_0} \right)
    \right]
\right) 
\right\} ,
\label{gperiod}
\end{eqnarray}
where $t=2 p L_0+z$, $p=0,1,2,...$, and $-L_0 \leq z \leq L_0$.
Given $t$, the value of the integer $p$ is obtained as
$p= \frac{1}{2} {\rm int}(t/L_0)$ or
$p= \frac{1}{2} [{\rm int}(t/L_0) + 1]$ for
${\rm int}(t/L_0)$ even or odd, respectively.
Note that during the
first period ($p=0$), $g(t)$ makes $R_1(t)$ vanish identically.
As we have already seen, since
the mirror is at rest
for $t<0$, we must impose $R(t)=t/L_0$ for $-L_0 \leq t \leq L_0$. Therefore
$a=A_n=B_n=0$, and the perturbative solution reads
\begin{equation}
R(t)={t\over L_0}+ \epsilon (-1)^{q+1} \left [ \frac{t}{L_0} 
\sin(\frac{q \pi t}{L_0})- \frac{z}{L_0} 
\sin(\frac{q \pi z}{L_0})\right ] .
\label{finalpert}
\end{equation}
The naive perturbative solution to the dynamical Casimir effect has 
been previously discussed in Ref. \cite{DodPLA}. However, in that
work the periodic function $g(t)$ was taken equal to zero. Thus, the solution
obtained there does not satisfy the correct boundary condition.

\section{RENORMALIZATION GROUP IMPROVEMENT}

We will now adapt the RG method of Section 2 in order to
obtain a solution to the Eq.(\ref{mooreq}) which is valid beyond the
short time  limit.
Let us introduce the arbitrary time $\tau$ and split $t$ as
$t-\tau +\tau$. The perturbative solution can then be written as
(see Eqs.(\ref{r0}) and (\ref{r1})),
\begin{eqnarray}
&& R(t) = a(\tau) + \sum_{n \geq 1} 
\left[A_n(\tau) \sin( \frac{n \pi t}{L_0} ) + 
      B_n(\tau) \cos( \frac{n \pi t}{L_0} ) \right] +
 \frac{t-\tau}{L_0} + 
\nonumber \\
&& \epsilon \frac{t-\tau}{L_0} (-1)^{q+1} 
\left(
\sin(\frac{q \pi t}{L_0}) + \frac{\pi}{2} \sum_{n \geq 1} n (-1)^n 
\left\{
A_n (\tau)\left[ \sin \left( \frac{(q+n) \pi t}{L_0} \right) +
 \sin \left( \frac{(q-n) \pi t}{L_0} \right)
    \right] +
\right. \right. \nonumber \\
&& \left. \left.
B_n (\tau) \left[ \cos \left( \frac{(q+n) \pi t}{L_0} \right) - 
\cos \left( \frac{(q-n) \pi t}{L_0} \right)
    \right] 
\right\}
\right) + g(t,\tau) + {\cal O}(\epsilon^2) ,
\label{Rgr}
\end{eqnarray}
where the bare parameters $a, A_n$ and $B_n$ have been replaced by their
renormalized counterparts $a(\tau), A_n(\tau)$ and $B_n(\tau)$.
Here $g(t,\tau)$ denotes the function $g(t)$ of Eq. (\ref{gperiod}) with
the same replacement. Note that then $g(t,\tau)$ is no more a periodic
function. 

The RG equation $(\partial R/\partial \tau)_t = 0$ consists 
in the present case 
of three
independent equations
\begin{eqnarray}
\frac{\partial a(\tau)}{\partial \tau} &=& \frac{1}{L_0}
+ {\cal O}(\epsilon^2) , \label{cons} \\
\frac{\partial A_n(\tau)}{\partial \tau} &=& 
   \epsilon \frac{(-1)^{q+1}}{L_0}
   \left[ \delta_{nq} + \frac{\pi}{2} \left\{ |n-q| A_{|n-q|} -
                                      (n+q) A_{n+q} \right\}
   \right] + {\cal O}(\epsilon^2) , \label{Aes} \\
\frac{\partial B_n(\tau)}{\partial \tau} &=& \epsilon
   \frac{\pi (-1)^{q+1}}{2 L_0} \left[|n-q| B_{|n-q|} + 
                                      (n+q) B_{n+q} \right]  
           + {\cal O}(\epsilon^2) , \label{Bes}
\end{eqnarray}
where we recall that the index $n$ is a positive integer. 
The solution to Eq.(\ref{cons}) is trivial: 
$a(\tau) = \frac{\tau}{L_0} + \kappa$, with $\kappa$ a constant to be
determined. If one writes
$A_n = \tilde{A}_n - \tilde{A}_{-n}$ and
 $B_n = \tilde{B}_n - \tilde{B}_{-n}$ ($n\geq 1$), where the new
variables satisfy
\begin{eqnarray}
\frac{\partial \tilde{A}_m}{\partial \tau^{\star}} &=& 
  \frac{2}{\pi} \delta_{mq} + (m-q)\tilde{A}_{m-q}-(m+q) \tilde{A}_{m+q}
+ {\cal O}(\epsilon^2) , 
\label{senos} \\
\frac{\partial \tilde{B}_m}{\partial \tau^{\star}} &=& (m-q) \tilde{B}_{m-q} 
+  (m+q) \tilde{B}_{m+q} + {\cal O}(\epsilon^2) , \label{cosenos}
\end{eqnarray}
then $A_n$ and $B_n$ satisfy Eqs.(\ref{Aes}) and (\ref{Bes})
respectively. 
Here we have introduced a new time
$\tau^{\star} \equiv \tau \epsilon \pi (-1)^{q+1}/(2 L_0)$.  
Since this set of first order differential
equations ensures the independence of the solution $R(t)$ with
 $\tau$, one can set $\tau=t$, which makes the terms proportional
to $\tau-t$ in Eq.(\ref{Rgr}) vanish identically. 

The initial conditions for these differential equations
are dictated by the perturbative solution: $a(0)=\tilde {A}_m(0)=
\tilde {B}_m(0)=0$. This means that $\kappa =0$ and that
$\tilde {B}_m (t)=0$ for all $t$. 
The coefficients $\tilde{A}_m$ are not all zero due to the presence of the
inhomogeneous term $\frac{2}{\pi} \delta_{mq}$. In order to solve the equation
corresponding to these coefficients we introduce the  generating 
functional $F(s,\tau^{\star}) = \sum_m s^m \tilde{A}_m(\tau^{\star})$.
Using Eq.(\ref{senos}) we see that it
satisfies the following differential equation
\begin{equation}
\frac{\partial F}{\partial \tau^{\star}} = \frac{2}{\pi} s^q +
\frac{\partial F}{\partial s} \left[ s^{q+1} - s^{1-q} \right] ,
\end{equation}
with boundary condition $F(s,\tau^{\star}=0)=0$. We make the following
Ansatz for the solution
$F(s,\tau^{\star}) = \Phi[e^{-\tau^{\star}} g(s)] + h(s)$, where 
$\Phi[\ldots]$, $g(s)$ and $h(s)$ are functions to be determined. Introducing
this form of the generating functional into the differential equation,
one determines the last two functions. The function $\Phi$ is determined
once the initial boundary condition is imposed. Finally the solution reads
\begin{equation}
F(s,\tau^{\star}) = - \frac{2}{\pi q} \ln \left[
\frac{e^{-q \tau^{\star}} (1+s^q) + e^{q \tau^{\star}} (1-s^q)}{2}
\right] .
\end{equation} 
In order to get the coefficients $\tilde{A}_m$ we expand this solution
in powers of $s$. In this way we obtain that the only non-vanishing
coefficients are 
$\tilde{A}_{m=0}(\tau^{\star})=-\frac{2}{\pi q} \ln(\cosh q \tau^{\star})$
and 
$\tilde{A}_{m=q j} = \frac{2}{\pi q j} \tanh^j(q \tau^{\star})$
with $j \in \Bbb N$. Note in particular that $\tilde{A}_{m<0} =0$,
which then means that the original coefficients $A_n$ are equal to the
$\tilde{A}_n$'s. 

The RG-improved solution for $R(t)$ can be obtained from
Eq.(\ref{Rgr}) by setting $\tau = t$. It is given by
\begin{equation}
R(t)= \frac{t}{L_0} + \sum_{j \geq 1} A_{qj}(t) 
\sin(\frac{q j \pi t}{L_0}) + \epsilon g(t,t) . 
\end{equation}
Using the explicit form of the coefficients $\tilde{A}_m$ we find
\begin {equation}
\sum_{j \geq 1} A_{qj}(t) 
\sin(\frac{q j \pi t}{L_0})=
- \frac{2}{\pi q} {\mbox{Im}} \ln \left[
1 + \xi + (1-\xi) e^{\frac{i q \pi t}{L_0}} \right] , \label{suma}
\end{equation}
where we have defined  
$\xi = \exp[\frac{ (-1)^{q+1} \pi q \epsilon t}{L_0}]$. The
(now non-periodic) function $g(t,t)$ can be easily evaluated
\begin{eqnarray}
g(t,t)&=&(-1)^q\frac{z}{L_0} \sin(\frac{q\pi z}{L_0})
\left[
1+\sum_{j\geq 1}A_{qj}(t)qj \cos(\frac{qj\pi z}{L_0})
\right] \nonumber\\
&=&(-1)^q\frac{z}{L_0} \sin(\frac{q\pi z}{L_0})
\left [ \frac{2\xi}{1+\xi^2 + (1-\xi^2)\cos(\frac {\pi q z}{L_0})}
\right ] .
\end{eqnarray}
Finally, the RG-improved solution reads
\begin{eqnarray}
R(t) &=& \frac {t}{L_0} - \frac{2}{\pi q} {\mbox{Im}} \ln \left[
1 + \xi + (1-\xi) e^{\frac{i q \pi t}{L_0}} \right] + \nonumber\\
&& \epsilon (-1)^q  \frac{z}{L_0} \sin(\frac{q\pi z}{L_0})
\left [ \frac{2\xi}{1+\xi^2 + (1-\xi^2)\cos(\frac {\pi q z}{L_0})}
\right ] .
\label{finalrg}
\end{eqnarray}
It is worth mentioning that this solution is valid as long as
 $\epsilon^2 \frac{t}{L_0}<1$, that is, the range of validity of the 
solution is longer than the perturbative one ($\epsilon \, \frac{t}{L_0} <1$).
In figure 1 we plot this function for the particular case $q=4$.

A solution to Eq.(\ref{mooreq}) in the long time limit was 
already obtained in \cite{DodJMP} using 
a different procedure. It coincides with our first two terms in
Eq. (\ref{finalrg}). There is perfect agreement between both solutions
at long times because it can be shown that the third term
in Eq. (\ref{finalrg}) is negligible in this limit (see Appendix A).
However, as we have mentioned in the previous Section, this term 
(that comes from the periodic function $g(t)$) is crucial for the solution 
to satisfy the correct boundary condition at short times.

\section{ENERGY DENSITY INSIDE THE CAVITY}

In order to study the local properties of the electromagnetic field inside
the cavity, we concentrate ourselves in the energy density of the field
\begin{equation}
\langle T_{00}(x,t) \rangle = \frac{1}{2} 
\left[
\langle \left( \frac{\partial A(x,t)}{\partial t} \right)^2 \rangle +
\langle \left( \frac{\partial A(x,t)}{\partial x} \right)^2 \rangle 
\right] ,
\label{TCERO}
\end{equation}
where the expectation values are taken with respect to the vacuum state.
Using the well-known point splitting method to regularize the divergence
appearing in the energy density \cite{FD}, one can obtain the following
expression for the renormalized energy density,
$\langle T_{00}(x,t) \rangle = -f(t+x)-f(t-x)$, where
\begin{equation}
f = \frac{1}{24 \pi} \left[
\frac{R'''}{R'} - \frac{3}{2} \left( \frac{R''}{R'} \right)^2 +
\frac{\pi^2}{2} (R')^2  \right] .
\label{EFE}
\end{equation}
This expression involves second and third derivatives of $R(t)$. As
 $R'(t)$ is discontinuous at $t=(2p+1) L_0$, $p=0,1,2,\ldots$ (see
Eq.(\ref{finalrg})), then the energy density will develop a delta function
singularity which will be infinitely reflected back and forth between the 
mirrors. The physical origin of this singularity is the initial discontinuity
of the wall velocity. We will ignore this singularity in what follows.

The structure of the electromagnetic field within the cavity for our
solution $R(t)$ is similar to that for other existing solutions
in the literature. In particular,
for $q \geq 2$ the energy density grows exponentially in the form
of $q$ travelling wave packets which become narrower and higher as time 
increases. The total energy within the plates increases 
exponentially at the expense of the energy needed to keep the plate moving. 
In figure 2 
we show the energy density profile between plates for a fixed time
and for the case $q=4$. As time
evolves, the peaks move back and forth bouncing against the mirrors. The 
height of the peaks increases as 
 $e^{\frac{2 \pi q \epsilon t}{L_0}}$ and their width decreases as
 $e^{- \frac{\pi q \epsilon t}{L_0}}$, so that the total area 
beneath each peak,
and hence the total energy, grows as $e^{\frac{\pi q \epsilon t}{L_0}}$. 
Apart from this exponential growth, there are `sub-Casimir' regions:
between the peaks the energy density takes values $q^2$ times smaller than the
static Casimir case, 
$\langle T_{00} \rangle_{\mbox{static}} = - \frac{\pi}{24 L_0^2}$. 
One can prove all
these properties analytically by computing the energy density with the
solution given in Eq. (\ref{finalrg}) and its derivatives (see Appendix B). 
In figure 3 the energy density is shown as a function of time at the mid point
between plates, also for the $q=4$ case. 

A rather different picture appears when one considers the $q=1$ case, that
corresponds to an oscillation frequency equal to the fundamental frequency
of the cavity. In this case the energy density at a given point oscillates in
time around the static Casimir value, and its time average coincides with
that value.

\section{Conclusions}

In this paper we have studied the one dimensional dynamical Casimir effect of
a resonant oscillating cavity. For this one dimensional case, the modes
of the electromagnetic field can be expressed in terms of the solution to the
so called Moore equation. We have used a renormalization group improvement of
the naive perturbative solution and we have succeeded in obtaining an
analytic solution which is valid up to times $t < L_0 \epsilon^{-2}$, thus
extending the range of validity of the perturbative solution 
 ($t < L_0 \epsilon^{-1}$). We have calculated 
the energy density inside the cavity and we have shown that a non trivial
structure appears, with a series of peaks that grow exponentially  in time and
move back and forth bouncing against the mirrors. Although this structure has
already been found in a previous work \cite{Cole}, here we have presented
an analytic derivation based on the renormalization group method described in
\cite{goldenfeld}.
We expect this
method to be useful to analyze the more realistic situation of a three 
dimensional oscillating cavity. This analysis can be performed by studying the
set of differential equations satisfied by the modes of the 
electromagnetic field \cite{DodPRA}. 
This topic will be the subject of further investigation.
   
\section{Acknowledgments}

We would like to thank Rafael Ferraro for his useful comments and C\'esar
Miquel for his technical assistance. D.A.R.D. thanks
Nicholas Phillips for useful discussions on related matters.
This research was supported by Universidad de Buenos Aires,
Consejo Nacional de Investigaciones Cient\'\i ficas y T\' ecnicas
and by Fundaci\' on Antorchas.

\appendix

\section{SHORT TIME AND LONG TIME BEHAVIOUR OF $\bbox{R(\lowercase{t})}$ }

In this Appendix we analyze the short time ($\epsilon \, \frac{t}{L_0} \ll 1$)
and long time ($\epsilon \, \frac{t}{L_0} \gg 1$) behaviour of the RG-improved
function $R(t)$ given in Eq.(\ref{finalrg}). Let us first split the solution
as $R(t)=R_{\rm s}(t) + R_{\rm np}(t)$, where

\begin{eqnarray}
R_{\rm s}(t) &=&  \frac {t}{L_0} - \frac{2}{\pi q} {\mbox{Im}} \ln \left[
1 + \xi + (1-\xi) e^{\frac{i q \pi t}{L_0}} \right] =
\frac{t}{L_0} - \frac{2}{\pi q} \arctan 
\left[ \frac{\sin(\frac{q \pi t}{L_0})}{\frac{1+\xi}{1-\xi}+ 
\cos(\frac{q \pi t}{L_0})} \right] ,
\label{Rrus} \\
R_{\rm np}(t) &=& \epsilon (-1)^q  \frac{z}{L_0} \sin(\frac{q\pi z}{L_0})
\left [ \frac{2\xi}{1+\xi^2 + (1-\xi^2)\cos(\frac {\pi q z}{L_0})}
\right ] ,
\label{Rnp}
\end{eqnarray}
with 
\begin{equation}
\xi = \exp \left[ \frac{(-1)^{q+1} \pi q \epsilon t}{L_0} \right].
\label{psi}
\end{equation}

The function $R_{\rm np}$ stems from the RG-improvement of the periodic
function $g(t)$, and it is non-periodic. The variable $z$ 
($-L_0 \le z \le L_0$) is given in terms of $t$ as $z=t-2 p L_0$ with
 $p=0,1,2,\ldots$. This integer $p$ is obtained from the
value that $t$ takes as $p=\frac{1}{2} {\mbox {int}}(t/L_0)$ or
 $p=\frac{1}{2} [{\mbox {int}}(t/L_0)+1]$, for ${\mbox {int}}(t/L_0)$ even or 
odd, respectively. 

For the short time limit $t \ll \epsilon^{-1} L_0$, these functions are

\begin{eqnarray}
R_{\rm s}(t) & \approx & \frac{t}{L_0} - (-1)^q \frac{\epsilon t}{L_0} 
\sin(\frac{q \pi t}{L_0}) \nonumber , \\
R_{\rm np}(t) & \approx & (-1)^q \frac{\epsilon z}{L_0}
\sin(\frac{q \pi z}{L_0}) ,
\end{eqnarray}
which then leads to the perturbative solution given in Eq.(\ref{finalpert}).

For the long time limit $t \gg \epsilon^{-1} L_0$ (but $t < \epsilon^{-2} L_0$
since this poses the upper limit for the validity of our RG solution),
we analyze $R_{\rm s}$ and $R_{\rm np}$ separately. 
We want to show that in this limit, the latter function is negligible. 
This can be graphically verified, but here we present 
an analytical demonstration. The function $R_{\rm s}$ has a first term, linear
in time, and a second one, that for late times becomes an oscillating function.
The amplitude of the oscillations is independent of $\epsilon$. Due to
this second term, $R_{\rm s}$ develops a staircase form for long times, as 
shown in Figure 1. Within regions of $t$ between odd multiples of $L_0$ (i.e.
in each period $p$), there appear $q$ jumps, located at values
of $t$ satisfying $\cos(\frac{q \pi t}{L_0})= \mp 1$, the upper sign
corresponding to even values of $q$ and the lower one to odd values of $q$.
Next we calculate the first derivative of $R_{\rm s}$. Since 
 $d \xi / dt$ is proportional to $\epsilon \xi$, one can differentiate the
function $R_{\rm s}$ with respect to time, treating $\xi$ as a constant.
The first derivative is then $R'_{\rm s}(t)=2 \xi \psi(t)$,
where 
\begin{equation}
\psi(t) = \frac{1}{L_0 \left[ 1+\xi^2+(1-\xi^2)\cos(\frac{q \pi t}{L_0})
\right]} .
\label{R1rus}
\end{equation}
Using Eq.(\ref{psi}), we see that $\xi$ vanishes (diverges) exponentially
for $q$ even (odd) at long times. For even $q$, the first derivative 
 $R'_{\rm s}$ develops peaks for times $t_n =\frac{(2n+1) L_0}{q}$, 
with $n$ an integer. 
The height of these peaks grows exponentially as $\xi^{-1}$. Between peaks,
 $R'_{\rm s}$ vanishes exponentially for long times.
In a similar fashion, for odd $q$, the first derivative develops peaks
at $t_n= \frac{2n L_0}{q}$.
In the following, we shall consider only the even case, the odd one being
completely similar.
 
Let us now analyze the function $R_{\rm np}$. Treating once again $\xi$ as
a constant when differentiating with respect to time, this function can be
expressed in terms of the first derivative of $R_{\rm s}$ as follows,
 
\begin{equation}
R_{\rm np}(t) = (-1)^q \epsilon z \sin(\frac{q \pi t}{L_0}) R'_{\rm s}(t) .
\label{R1np}
\end{equation}
We see that $R_{\rm np}$ is the product of a bounded factor ($|z|<L_0$)
times a function $F(t) \equiv \sin(\frac{q \pi t}{L_0}) R'_{\rm s}(t)$ that is
proportional to the first derivative of $R_{\rm s}$ and might thus be 
unbounded. We shall now show that this is {\it not} the case. 
Far from the position of the peaks, $F$ is bounded because $R'_{\rm s}$ is.
In a surrounding of $t_n$, we express $F$ as

\begin{equation}
F(\delta) = - \frac{2 \pi q \xi \delta}
{L_0 \left[ (1-\xi^2) \frac{\pi^2 q^2 \delta^2}{2} + 2 \xi^2 \right]}
\end{equation}
where $\delta=(t-t_n)/L_0$.
Firstly, we note that this function vanishes for $\delta=0$, i.e. 
 $R_{\rm np}(t=t_n) = 0$. Secondly, this function has 
extrema equidistant from $t_n$ located at 
 $\delta_{\pm} = \pm \frac{2 \xi}{\pi q \sqrt{1-\xi^2}}$ and at these points 
 $F(\delta_{\pm}) = \mp \frac{1}{L_0 \sqrt{1-\xi^2}}$. Since for long times
 $\xi \rightarrow 0$, we conclude that $F(\delta_{\pm})$ is bounded by
 $1/L_0$. Consequently, $R_{\rm np}$ is a correction of order $\epsilon$ to
the second term of Eq.(\ref{Rrus}).

\section{STRUCTURE OF THE ELECTROMAGNETIC FIELD} 

In this Appendix we study briefly the structure of the electromagnetic
field within the cavity as given by Eqs.(\ref{TCERO},\ref{EFE}) 
for our solution
 $R(t)$, in order to understand the form of the
energy density profile shown in figures 2 and 3 in the long time regime
 $\frac{\epsilon t}{L_0} > 1$. As in Appendix A we 
will split $R(t)$ as $R(t)=R_{\rm s} + R_{\rm np}$ (see 
Eqs.(\ref{Rrus},\ref{Rnp})). 
In order to analyze the energy-momentum tensor, we need to study the first 
three derivatives of the solution $R(t)$. For $R_{\rm s}$ we get

\begin{eqnarray}
R'_{\rm s}(t) &=& 2 \xi \psi(t) \label{runo},  \\
R''_{\rm s}(t) &=& 2 \xi (1-\xi^2) \pi q \sin(\frac{\pi q t}{L_0}) 
\psi^2(t) \label{rdos}, \\
R'''_{\rm s}(t) &=& 2 \xi (1-\xi^2) (\pi q)^2 [(1+\psi^2) 
\cos(\frac{\pi q t}{L_0}) + (1-\xi^2) [1+\sin^2(\frac{\pi q t}{L_0})]] 
\psi^3(t) \label{rtres}\\
\psi(t) &=& \frac{1}{L_0 \left[ 1+\xi^2+(1-\xi^2)\cos(\frac{q \pi t}{L_0})
\right]} ,
\end{eqnarray}
where again we treat $\xi$ as a constant.
For even $q$, all these derivatives develop peaks at times
 $t_n =\frac{(2n+1) L_0}{q}$. As can be easily seen from 
the above equations, the 
height of the peaks for the m-th derivative of $R_{\rm s}$ is proportional 
to $\xi^{-m}$.
Using the same methods as in Appendix A, one can show (after some algebra)
that near the times $t_n$, the m-th derivative of $R_{\rm np}$ also has peaks
whose heights are proportional to $\epsilon \xi^{-m}$. Since 
 $\epsilon \ll 1$, it means that at long times all the derivatives of 
 $R_{\rm np}$ are negligible with respect to those of $R_{\rm s}$. The 
form of the energy-momentum tensor will be governed just by
the first part of our solution, namely by $R_{\rm s}$. 

Let us concentrate only on the contribution to the energy density which is 
proportional to $R'^2$ (see Eq. (\ref{EFE})). From the above discussion it is
clear that $T_{00}$ will develop peaks which grow as 
 $e^{\frac{2 \pi q \epsilon t}{L_0}}$.
Their width decreases exponentially as 
 $e^{- \frac{\pi q \epsilon t}{L_0}}$, so the total 
area of the peaks grows exponentially. The same holds for the total energy in 
the cavity. The analysis of the other two terms of Eq.(\ref{EFE}) lead to the 
same conclusion.

The case $q=1$ shows a different behaviour. Indeed, when the energy density of
the field is computed using the derivatives of $R_{\rm s}$ given above,
there is a cancellation between the different contributions in Eq.(\ref{EFE})
and the final answer coincides with the static Casimir value.

%
%

\begin{figure}[h]
\centering \leavevmode
\epsfxsize=12cm
\epsfbox{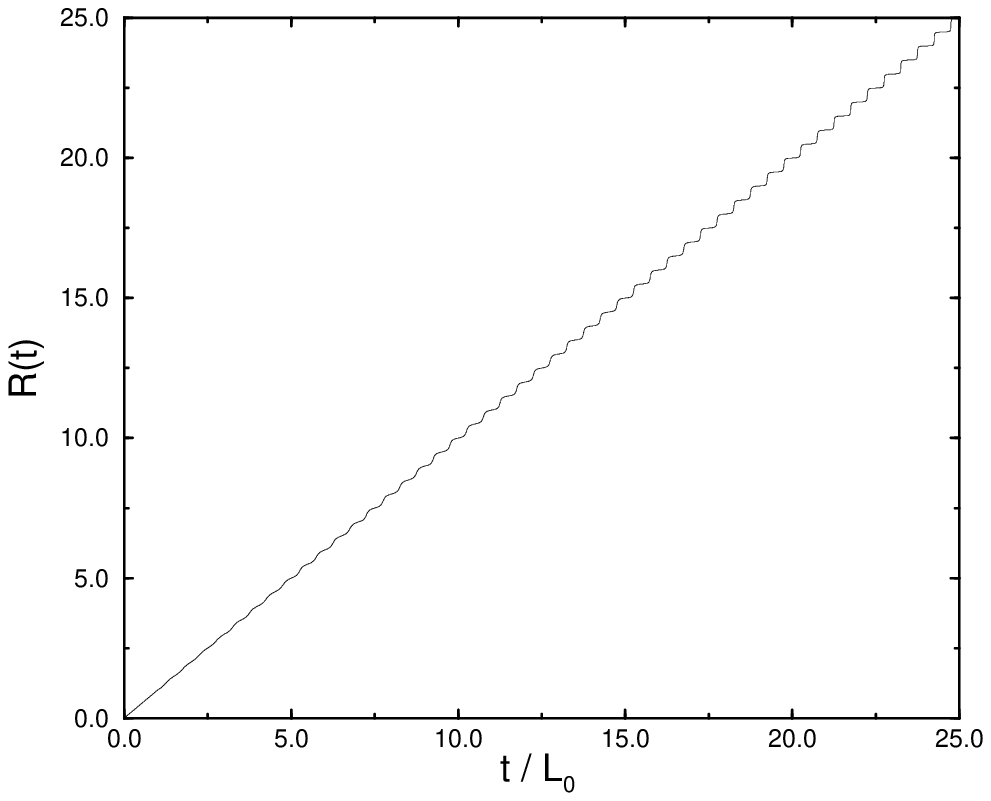}
\caption{$R(t)$ vs. $t/L_0$ as given by Eq.(\ref{finalrg}). 
The values of the parameters are $q=4$ and $\epsilon=0.01$.}
\end{figure}

\begin{figure}[h]
\centering \leavevmode
\epsfxsize=12cm
\epsfbox{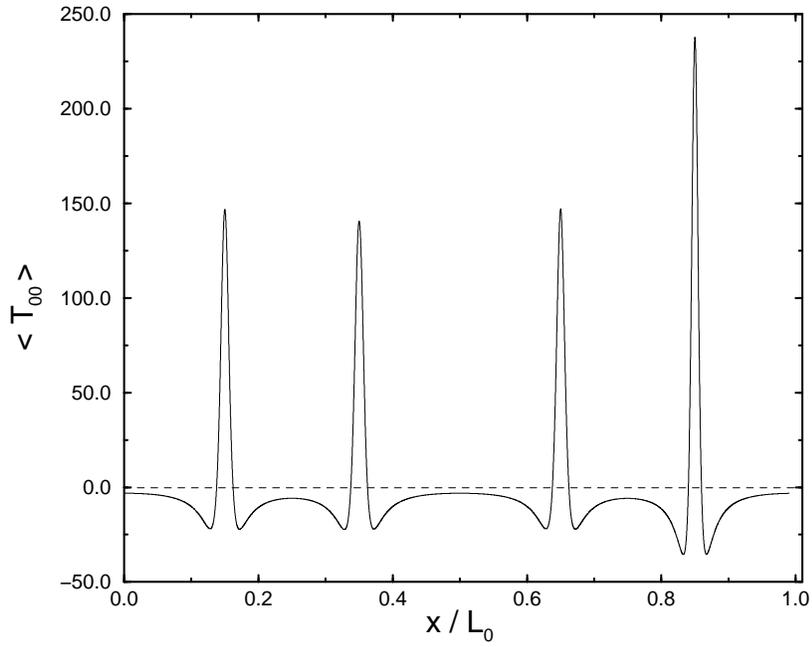}
\caption{Energy density profile between plates for fixed time $t/L_0=20.4$ for
the $q=4$ case. The amplitude coefficient is $\epsilon=0.01$.}
\end{figure}

\begin{figure}[h]
\centering \leavevmode
\epsfxsize=12cm
\epsfbox{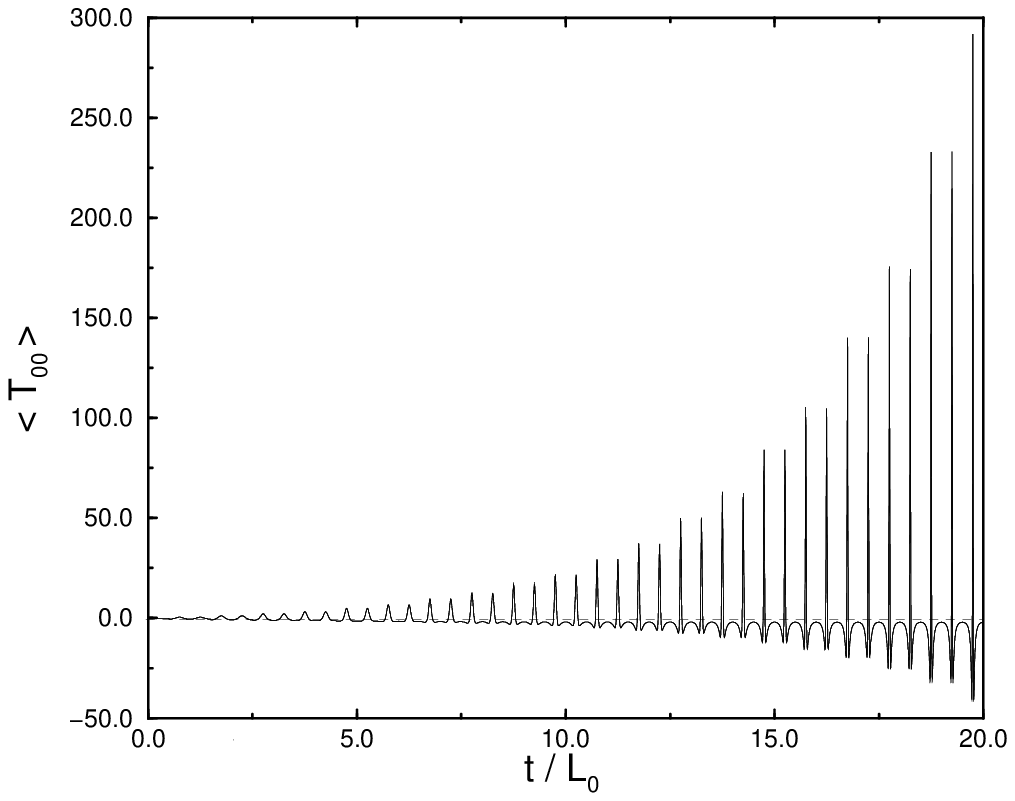}
\caption{Energy density as a function of time for the mid-point $x/L_0=0.5$
between plates. The parameters are $q=4$ and $\epsilon=0.01$. }
\end{figure}


\begin{references}

\bibitem{Casimir} H. B. G. Casimir, Proc.\ K.\ Ned.\ Akad.\ Wet.\ 
{\bf 51}, 793 (1948).

\bibitem{Moore} G. T. Moore, J.\ Math.\ Phys.\ {\bf 11}, 2679 (1970).

\bibitem{FD} S. A. Fulling and C. W. Davies, Proc.\ R.\ Soc.\ Lond.\ A 
{\bf 348}, 393 (1976).

\bibitem{Rafael} M. Castagnino and R. Ferraro, Ann.\ Phys.(N.Y.)
{\bf 154}, 1 (1984).

\bibitem{Law} C. K. Law, Phys.\ Rev.\ Lett.\ {\bf 73}, 1931 (1994).

\bibitem{Cole} C. K. Cole and W. C. Schieve, Phys. Rev. A {\bf 52}, 
4405 (1995).

\bibitem{DodPLA} V. V. Dodonov et {\it al.}, Phys.\ Lett.\ A {\bf 149},
225 (1990).

\bibitem{DodJMP} V. V. Dodonov et {\it al.}, J.\  Math.\ Phys.\ {\bf 34}, 
2742 (1993).

\bibitem{DodPRA} V. V. Dodonov and A. B. Klimov, Phys.\ Rev.\ A {\bf 53}, 
2664 (1996).

\bibitem{goldenfeld} L. Y. Chen et {\it al.}, Phys.\ Rev.\ Lett.\ {\bf 73},
1311 (1994); Phys.\ Rev.\ E {\bf 54}, 376 (1996). 

\bibitem{bender} C. M. Bender and S. A. Orszag, {\it Advanced Mathematical
Methods for Scientists and Engineers} (McGraw-Hill, New York, 1978).

\end{references}
\end{document}